\documentclass[12pt]{article}
\usepackage{amsmath,mathrsfs,amsthm,amsfonts}
\usepackage{graphicx}
\usepackage{enumerate}
\usepackage{natbib}
\usepackage{url} 
\usepackage{xcolor}
\usepackage{cleveref}
\usepackage{bm}
\usepackage{tikz}
\usepackage{booktabs}
\usepackage{threeparttable}
\usetikzlibrary{trees}
\usetikzlibrary{shapes,decorations,arrows,calc,arrows.meta,fit,positioning}
\usetikzlibrary{shapes.multipart}
\tikzset{
	-Latex,auto,node distance =1 cm and 1 cm,semithick,
	state/.style ={ellipse, draw, minimum width = 0.7 cm},
	state1/.style ={ draw, minimum width = 0.7 cm},
	point/.style = {circle, draw, inner sep=0.04cm,fill,node contents={}},
	bidirected/.style={Latex-Latex,dashed},
	el/.style = {inner sep=2pt, align=left, sloped}
}

\newcommand{\blind}{1}

\renewcommand{\b}[1]{\boldsymbol{#1}}
\definecolor{dblue}{HTML}{0072B2}
\definecolor{dorange}{HTML}{D55E00}
\definecolor{dgreen}{rgb}{0.,0.6,0.}

\crefname{equation}{}{}

\crefname{theorem}{Theorem}{Theorems}

\crefname{lemma}{Lemma}{Lemmas}

\crefname{requirement}{Consideration}{Considerations}

\newtheorem{proposition}{Proposition}
\newtheorem{assumption}{Assumption}

\begin{document}

\def\spacingset#1{\renewcommand{\baselinestretch}%
  {#1}\small\normalsize} \spacingset{1}


\if1\blind
{
 \begin{center}
 	\spacingset{1.5} 
	{\Large\bf  Nonparametric identification of causal effects in clustered observational studies with differential selection} \\ \bigskip \bigskip
	{\large  Ting Ye\footnote{University of Washington, Email: tingye1@uw.edu}, Ted Westling\footnote{University of Massachusetts Amherst, Email: twestling@umass.edu},
		Lindsay Page\footnote{Brown University, Email: lindsay\_page@brown.edu},  and Luke Keele \footnote{University of Pennsylvania, Email: luke.keele@gmail.com.}}
  \end{center}

} \fi

\if0\blind
{
  \bigskip
  \bigskip
  \bigskip
  \begin{center}
  	\spacingset{1.5} 
  {\Large\bf  Nonparametric identification of causal effects in clustered observational studies with differential selection}
  \end{center}
  \medskip
} \fi

\bigskip

\begin{abstract}
{
	The clustered observational study (COS) design is the observational study counterpart to the clustered randomized trial. In a COS, a treatment is assigned to intact groups, and all units within the group are exposed to the treatment. However, the treatment is non-randomly assigned. COSs are common in both education and health services research. In education, treatments may be given to all students within some schools but withheld from all students in other schools. In health studies, treatments may be applied to clusters such as hospitals or groups of patients treated by the same physician. In this manuscript, we study the identification of causal effects in clustered observational study designs. We focus on the prospect of differential selection of units to clusters, which occurs when the units' cluster selections depend on the clusters' treatment assignments. Extant work on COSs has made an implicit assumption that rules out the presence of differential selection. We derive the identification results for designs with differential selection and that contexts with differential cluster selection require different adjustment sets than standard designs. We outline estimators for designs with and without differential selection. Using a series of simulations, we outline the magnitude of the bias that can occur with differential selection. We then present two empirical applications focusing on the likelihood of differential selection.

}

\end{abstract}

\noindent%
{\it Keywords:} Causal inference,
Identification, Clustered Observational Study
\vfill

\newpage
\spacingset{1.5} 

\section{Introduction}

Many applied analyses focus on whether a treatment given to some set of units causes a hypothesized effect. In some settings, treatments of interest are allocated individually. That is, a treatment is assigned to certain people individually and not to others. However, in other settings, the treatment is allocated to groups or intact clusters of units -- e.g., to hospitals or schools -- while outcomes of interest are measured at the unit level--- e.g., patients or students. 	The critical feature of such treatment assignment processes is that all or none of the units within a cluster are exposed to the treatment of interest. When group-level treatments are randomly assigned, the study design commonly is referred to as a clustered randomized trial (CRT) \citep{raudenbush1997statistical, hedges2007intraclass}. In a clustered observational study (COS), treatment is still assigned at cluster level but assignment is non-random \citep{pagedesign2019}. Given non-random assignment in a COS, differences in outcomes may reflect pretreatment differences in treated and control groups rather than actual treatment effects \citep{hansen2014clustered}. Moreover, the COS design requires specialized forms of statistical adjustment for observed confounders \citep{Keele:2015,Keele:2016b}. Next, we highlight two areas of applied research where the COS design is common.

\subsection{Education}

COSs are common in educational research, where treatments are often applied to schools. For example, \citet{adelson2012examining} study how changes in gifted programs in some schools affect student-level outcomes.  Much research has focused on whether Catholic schools are more effective than public schools in fostering student achievement \citep{Coleman:1982,Hoffer:1985,Coleman:1987}. Often a new reading program may be implemented in some schools but not in others \citep{pagedesign2019}. Other interventions broadly seek to change entire school structures in hopes of improving chronically underperforming schools \citep{bryk2015, mehta2012, mcguinn2016}. For example, in 2015-16, the Wake County Public School System implemented a school turnaround strategy known as the Elementary Support Model (ESM) in 12 selected schools. Schools in the ESM condition received a range of supports over three years, including governance reform, additional staffing, and instructional coaching \citep{paeplow2019}. This program implementation is prototypical of the COS template; the intervention is non-randomly assigned to and delivered at the school level, but the investigators are focused on academic and behavioral outcomes measured at the student level.

\subsection{Health Services Research}

The COS design is also common in comparative effectiveness research (CER), which focuses on evaluating the causal effects of health care strategies on patient outcomes \citep{hernan2018c}. CER encompasses patient level treatments and the effects health care delivery, organization and financing, as well as public health interventions \citep{iom2009}. In CER, COS designs often mirror those in education where interventions are applied to entire hospitals. For example, the COS design has been applied at the hospital level to study the effect of the work environment and amount of autonomy given to nurses \citep{Rao:2017dp,Silber:2016qe}, the effect of medical residents' duty hours \citep{Bilimoria:2016la,Patel:2014ij,Silber:2014kl}, and the effect of Magnet certification for high quality nursing \citep{Barnes:2016dz,McHugh:2013fy}. The COS design in CER also arises at the level of the physician. In this setting, a treatment is applied to some physicians but not others, but outcomes are measured at the patient level. COS designs of this type include studies on the effects of training in a teaching hospital \citep{Navathe:2013th,Navathe:2013bs,Srinivas:2013eu,Lorch:2012fv} and the effect of a university-based surgical residency \citep{sellers2018association}.

\subsection{Clustered Observational Studies}

In both of these applied fields, a randomized trial would require randomly assigning entire clusters to treatment or control status. However, in educational and health settings, such randomization can often be impractical or even impossible, and so we require different analytic strategies to address questions of cause and effect. In such cases, critical research questions can only be investigated using clustered observational studies. The extant literature on the COS study design, however, critically assumes that the population of units within the clusters is not affected by treatment assignment \citep{Keele:2015,Keele:2016b,hansen2014clustered}. Here, we focus on a feature of clustered treatment assignment that may result in a form of selection bias arising from differential selection of units -- patients or students -- across treated clusters. Specifically, we consider the possibility that the population of units within treated clusters changes in reaction to the treatment being assigned to the cluster. For example, once an intervention such as ESM is put into place, families with high achieving students may move to treated school catchment areas. Under this form of differential selection, differences in outcomes between treated and control students with similar pre-treatment characteristics may be a result of peer effects rather than the treatment effect, because schools with ESM may attract higher achieving students. That is, ESM may appear to have improved outcomes, when in fact it only attracted higher achieving students.
\citet{ogburn2014causal} refer to this as allocational interference. 

In this manuscript, we present a series of new results for clustered observational studies. First, we develop a notational framework that allows for differential selection of units to clusters as a function of treatment assignment. We use the target trial framework to define three distinct trials that differ in terms of whether unit to cluster assignment is affected by the treatment. We highlight how extant COS research has been based on a hidden assumption that rules out the possibility of differential selection. We show that identification for each target trial depends on conditioning on different types of covariates, to which we refer as the conditioning or adjustment set.  We also outline the additional assumption that identifies causal effects under differential selection. We then conduct a simulation study investigating how bias can arise in the analysis of COS designs, when the wrong target trial is assumed. We present results from two different applications, one from education and one from health services research. Finally, we conclude with a discussion of key implications for applied research.

\section{COS Framework}

To understand issues of causal identification in this context, we use the target trial framework. Target trial emulation calls for applying design principles from randomized trials to the analysis of observational data \citep{Hernan:2016}. More specifically, in the target trial framework, the investigator derives estimands and identification conditions from the hypothetical experimental trial that is being emulated. From a practical standpoint, the target trial of interest may be infeasible as an actual randomized trial. However, that is largely irrelevant for our technical, analytic purpose, which is to use the target trial to structure the design and analysis of an observational study. Next, we outline two hypothetical target trials that are relevant to COS designs.

\subsection{Target Trials for the COS Design}

\begin{figure}[h]
	\centering
	\makebox{\includegraphics[width=0.9\textwidth]{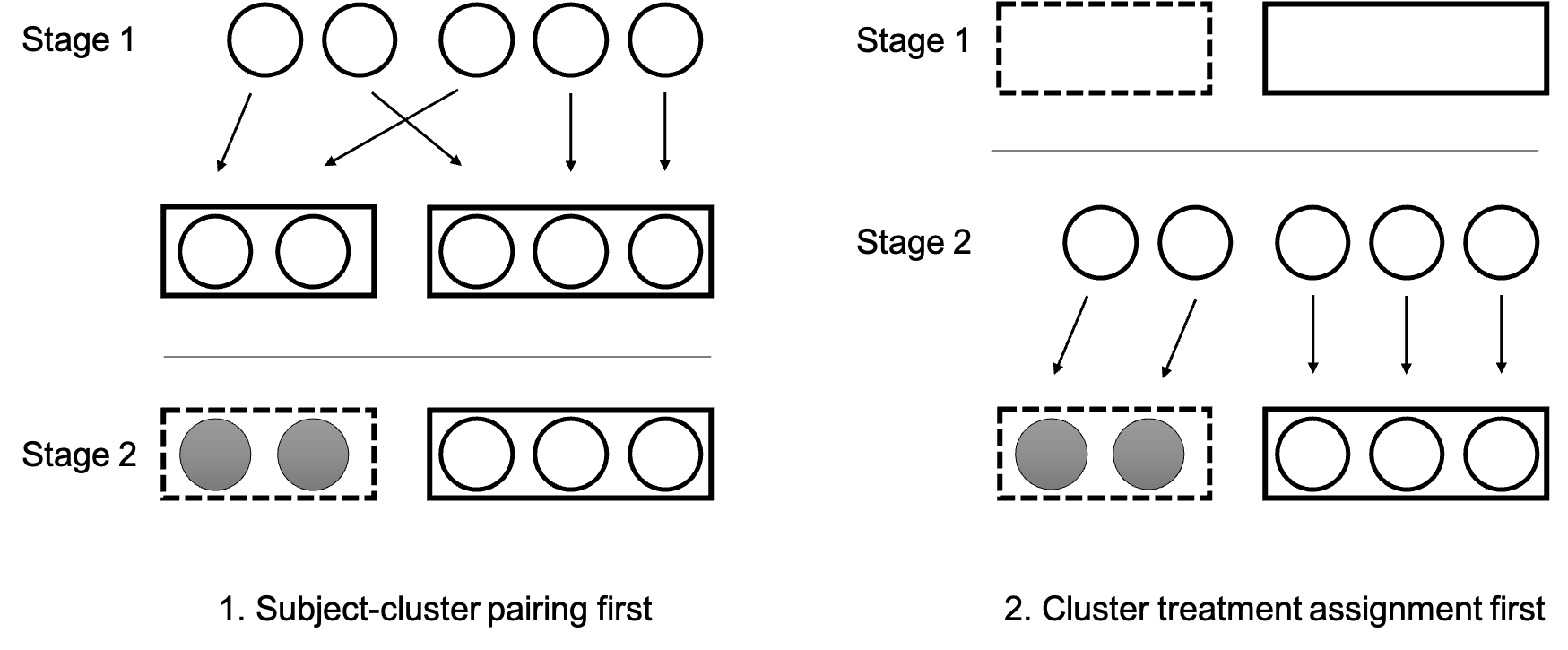}}
	\caption{\label{fig:schematic} Schematic of two different types of target trials with different unit-cluster selection mechanisms.  Dashed line represents cluster-level treatment assignment. The gray color represents treated units.}
\end{figure}

Here, we provide an informal outline of two hypothetical target trials, where in both cases there are $ n $ units and $ m $ clusters. In both of these target trials, there are two stages of assignment, the order of which represent the key difference between these two target trials. In Figure~\ref{fig:schematic}, we provide a graphical illustration of these two target trials, with individual units represented by circles and groups or clusters represented by rectangles. In target trial 1, units are first assigned (randomly or not) to clusters and then clusters are randomly assigned to treatment or control. Target trial 1 also includes the setting where the unit-cluster pairing at the first stage is fixed, which is typical in the literature on CRTs. In target trial 2, on the other hand, we consider a scenario where at the first stage, $ m $ clusters are randomized to receive treatment or control, and at the second stage, $ n $ units are randomly assigned to the $ m $ clusters. This setting is also common in CRTs when units are recruited after clusters are randomized. 

If the two stages of randomization are independent of each other---if the  stage 2 randomization does not depend on the outcome of stage 1 assignments---we will show that the identification conditions for target trial 1 and 2 are equivalent. However, these two target trials differ if the treatment assignment depends on unit-cluster pairing in target trial 1, or if the unit-cluster pairing depends on clusters' treatment assignments in target trial 2.  In particular, target trial 1 precludes the possibility of differential selection, but target trial 2 does not when unit selection is no longer randomized at the second stage. Next, we develop notation that allows us to formally describe these two target trials.
	
\subsection{Notation}

Let ${\b{A}} = (  A_1, \dotsc,  A_m)$ be the observed binary treatment indicators for all clusters, and let $\b{J}({\b{a}} ) = (J_1(\b{{\b{a}} }), \dots, J_n(\b{{\b{a}} }))$ denote the clusters that the $ n $ subjects would be assigned to had the clusters received exposure: ${\b{a}} = ( a_1, \dotsc,  a_m)$. We use this potential-selections notation to capture the fact that the subject-cluster pairing may depend on treatment status of clusters, ${\b{A}}$. We use $ \b{J} = \b{J} (\b{A})$ to denote the observed cluster values for the $ n $ subjects. Next, let $Y_i({\b{a}}, \b{j})$ be the potential outcome for subject $ i $   had $ {\b{A} } = {\b{a}}$ and $ \b{J}=  \b{j} $, where $ \b{j} = (j_1,\dots, j_n) $. This notation allows for unit outcomes to depend on both treatment \emph{and} cluster assignments.  Finally, $Y_i = Y_i(\b{A}, \b{J})$ is the observed outcome for subject $ i $.

For each unit $ i $, we observe a vector of baseline covariates  $ X_i $ that describe the units (age, race, etc), and for each cluster $ j $, we also observe baseline covariates $ W_j $ that describe cluster characteristics (school or hospital size). Denote  $ \b{W} = (W_1,\dots, W_m) $, $ \b{X} = (X_1,\dots, X_n) $, and $ \b{Y}  =(Y_1,\dots, Y_n) $. Given the natural clustering in a COS, we also observe covariates that are aggregates of the individual-level covariates in the cluster denoted as $ h( \b{X}_{[j]} )$, where $\b{X}_{[j]} = \{ X_1,\dots, X_n: J_i =j  \}$  is the collection of covariates for subjects belonging to the cluster $j$, which is a function of  $(\b{X}, \b{J}) $.  Some commonly-used  $h(\cdot)$ functions include the mean and quantile functions.

We assume that subjects' potential outcomes depend on $ \b{A} $ only through the exposure value of their own cluster, but not other clusters' exposure values, i.e., for any two exposure vectors $ \b{a} $ and $ \b{a}' $ such that  $ \b{a}_{{j}_i}=   \b{a}_{{j}_i}' $, we have  $ Y_i(\b{a}, \b{j} ) =   Y_i(\b{a}', \b{j} ) $. Thus, the potential outcome $ Y_i({\b{a}}, \b{j})$  can be simplified and written as  $ Y_i( {a}_{{j}_i}, \b{j})$, and the observed outcome satisfies $ Y_i = Y_i(\b{A}, \b{J})= Y_i({A}_{{J}_i}, \b{J}  )$. In the identification analysis that follows we focus on a common target estimand the average treatment effect formalized as:
\begin{equation}
\label{eq:estmd}
E[Y_i(1, \b{J}) - Y_i(0, \b{J}) ].
\end{equation}
\noindent This estimand represents the average difference between the potential outcomes that would be observed if all clusters were treated versus if all clusters were untreated, while the cluster membership is kept as the observed (possibly random) value in the actual observational study, which for the second trial could depend on the observed treatment. In target trial 2, this estimand is similar to that of a natural direct effect in a mediation analysis \citep{vanderweele2016mediation}.

\section{Identification}
\label{sec: identification}

Next, we turn to issues of causal identification. Under the target trials articulated in Figure \ref{fig:schematic} with randomization of both units and clusters, the average treatment effect is identified.  Here, we derive the relevant identification conditions for both of the target trials reconstituted as observational studies assuming treatments are not randomly assigned. We focus on the identification conditions through conditioning on baseline covariates. We outline that each target trial identification depends on different adjustment or conditioning sets. We demonstrate, in particular, that the presence of differential selection at the individual level requires conditioning on unit-level covariates.

\subsection{Target Trial 1: Fixed Unit-Cluster Pairing}

First, we focus on target trial 1 where unit-cluster pairing occurs before treatment assignment. For target trial 1, identification of the treatment effect is possible using a version of the standard conditional ignorability assumption altered to reflect cluster-level treatment assignment \citep{vanderweele2008ignorability,hansen2014clustered}. That is, one assumes that treatment assignment is random within strata of the cluster-level covariates $W_j$ and $ h( \b{X}_{[j]} )$ which are aggregates of the individual-level covariates in the cluster. Formally, the key identification assumption is written as
\begin{assumption}[Target Trial 1] \label{assump: pairing first}
	(i) $\b{J}= \b{J} (1) = \b{J} (0)$; (ii) For every $i, j$, $a$, and $\b{j}$, $A_j \perp \{ \b{J}, \b{X},Y_i( a, \b{j}) \} \mid W_j ,  h( \b{X}_{[j]} )$.
\end{assumption}

Assumption~\ref{assump: pairing first}(i) describes a unit-cluster pairing mechanism that is unaffected by cluster-level treatment assignments. Assumption~\ref{assump: pairing first}(ii) says that each cluster's treatment assignment probability is a function of the cluster's characteristics and  certain aggregate of the individual-level covariates in the cluster. We should note that outside of the COS setting, it is not common to require $\b{X}$ conditionally independent of $A_j$.  Figure~\ref{fig1} represents this assumption for each cluster as a directed acyclic graph (DAG). The key feature of this DAG is that conditioning on $W_j$ and $ h( \b{X}_{[j]} )$ blocks all backdoor paths from $A_j$ to $\b{Y}$ and renders the treatment-outcome relationship unconfounded \citep{pearl1995causal}. 

\begin{figure}
	\centering
	\begin{tikzpicture}
		\node[state] (a) at (0,0) {${A}_j$};
		\node[state] (y) [right =of a, xshift=1cm] {$\b{Y}$};
		\node[state] (w) [rectangle, right =of a, yshift=1cm, xshift=-0.6cm] {$W_j$};
		\node[state] (h) [rectangle, below =of a] {$ h( \b{X}_{[j]} )$};
		\node[state] (j) [right = of h, xshift=0.9cm] {$\b{J}$};
		\node[state] (x) [right =of h, yshift=0.8cm, xshift=-0.5cm] {$\b{X}$};
		\path (a) edge node[above]{} (y);
		\path (w) edge node[above]{} (a);
		\path (x) edge node[above]{} (h);
		\path (j) edge node[above]{} (h);
		\path (h) edge node[above]{} (a);
		\path (w) edge  node[above]{} (y);
		\path (j) edge node[above]{} (y);
		\path (x) edge node[above]{} (y);
		\path (x) edge node[above]{} (j);
		\path (w) edge node[above]{} (j);
	\end{tikzpicture}
	\caption{A DAG illustrating identification conditions for cluster $j$ in target trial 1. Conditioning on $W_j$ and $ h( \b{X}_{[j]} )$ renders the treatment-outcome relationship unconfounded.}
	\label{fig1}
\end{figure}

Next, we provide a formal statement of identification under this assumption. Define $\mu_{\rm wh} (a,w, h)=  E[ Y_i \mid  A_{J_i }=a, W_{J_i}=w, h(  \b{X}_{[J_i]} ) = h] $ and $ \tau_{\rm wh}=	E[ \mu_{\rm wh}(1, W_{J_i},   h(  \b{X}_{[J_i]} ) ] - E[ \mu_{\rm wh}(0, W_{J_i},   h(  \b{X}_{[J_i]} ) ] $. Similarly, define  $\mu_{\rm whx}(a, w, h, x)= E[ Y_i \mid  A_{J_i }=a, W_{J_i}=w,
h(  \b{X}_{[J_i]} ) = h , X_i = x ] $ and  $ \tau_{\rm whx}=	E[ \mu_{\rm whx}(1, W_{J_i}, h(  \b{X}_{[J_i]} ),  X_i) ] - E[ \mu_{\rm whx}(0, W_{J_i},  h(  \b{X}_{[J_i]} ),  X_i) ] $.

Proposition \ref{prop: a} shows that there are two ways of identifying the treatment effect in target trial 1.
\begin{proposition}[Target Trial 1]\label{prop: a}
	Under Assumption \ref{assump: pairing first},  $\tau_{\rm wh} = \tau_{\rm whx} = E [ Y_i (1, \b{J}) ] - E [ Y_i (0, \b{J}) ]$.
\end{proposition}
The proof of Proposition \ref{prop: a} and all other proofs are in the supplementary material. Proposition \ref{prop: a} shows that the treatment effect is identifiable even if we do not condition on unit-level covariates $\b{X}$. This is due to the fact that in Assumption~\ref{assump: pairing first}, $\b{X}$ does not directly affect clusters' treatment assignment. However, adjusting for unit-level covariates may improve efficiency.

\subsection{Target Trial 2: Cluster Treatment Assignment First}

Next, we consider identification for target trial 2. Here, the identification conditions differ from target trial 1. Critically, we must now account for the possibility that when cluster treatment assignment precedes unit-cluster pairing, $\b{J}$ is \emph{post-treatment} and can directly affect subjects' outcomes. Notably, under target trial 2, we consider two different mechanisms for how subjects are selected into clusters. As such, we split target trial 2 into two target trials that we denote as 2(a) and 2(b). Specifically, we denote blinded unit-cluster pairing as target trial 2(a), and we denote unblinded unit-cluster pairing as target trial 2(b). Here, differential selection is possible under target trial 2(b), due to the unblinded unit-cluster pairing mechanism.

\subsubsection{Target Trial 2(a): Blinded Unit-Cluster Pairing}

For target trial 2(a), units
are blinded to clusters' treatment status when selecting clusters, which we formalize through the following assumption:
\begin{assumption}[Target Trial 2] \label{assump: blinded} (i) $\b{J}= \b{J} (1) = \b{J} (0)$; (ii) for every $i,j, a,$ and $ \b{j}$, $	A_j \perp \{ \b{J}, \b{X},Y_i( a, \b{j}) \} \mid W_j $.
\end{assumption}
Assumption \ref{assump: blinded} formalizes the scenario where at the first stage, clusters (e.g., schools, hospitals, physicians) adopt the treatment or not independently, and at the second stage, subjects select clusters with knowledge of clusters' characteristics $\b{W}$ but with no knowledge of clusters' treatment assignment $\b{A}$. The key difference under Assumption \ref{assump: blinded} we are no longer conditioning on $ h( \b{X}_{[j]} )$. As such, for target trial 2(a) identification is possible by conditioning on reduced set of covariates as compared to target trial 1.

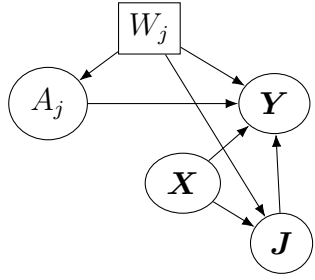
\begin{figure}
	\centering
	\centering
	\begin{tikzpicture}
		\node[state] (a) at (0,0) {${A}_j$};
		\node[state] (y) [right =of a, xshift=1cm] {$\b{Y}$};
		\node[state] (w) [rectangle, right =of a, yshift=1cm, xshift=-0.6cm] {$W_j$};
		\node[state] (j) [right = of h, xshift=0.9cm] {$\b{J}$};
		\node[state] (x) [right =of h, yshift=0.8cm, xshift=-0.5cm] {$\b{X}$};
		\path (a) edge node[above]{} (y);
		\path (w) edge node[above]{} (a);
		\path (w) edge  node[above]{} (y);
		\path (j) edge node[above]{} (y);
		\path (x) edge node[above]{} (y);
		\path (x) edge node[above]{} (j);
		\path (w) edge node[above]{} (j);
	\end{tikzpicture}
	\caption{ A DAG illustrating identification conditions for cluster $j$ in  target trial 2(a). Conditioning on $W_j$ renders the treatment-outcome relationship unconfounded.} \label{fig2}
\end{figure}

This scheme is illustrated by a DAG in Figure~\ref{fig2}, which  would arise, for  example, when  a new reading curriculum is assigned to some schools but not others. If students and their parents are unaware of the new reading curriculum when deciding which school to attend, this would preclude the possibility that the student population could shift to different schools in response to the treatment. On the other hand, it does allow for a scenario, where parents select schools based on school characteristics such as test scores history or lagged student demographics.

Next, we formalize identification under Assumption \ref{assump: blinded}. Define $\mu_{\rm w} (a,w)=  E[ Y_i \mid  A_{J_i }=a, W_{J_i}=w] $ and $ \tau_{\rm w}=	E[ \mu_{\rm w}(1, W_{J_i} ) ] - E[ \mu_{\rm w}(0, W_{J_i}) ] $.  Under Assumption \ref{assump: blinded},  we can identify the average treatment effect in three different ways.
\begin{proposition}[Target Trial 2(a)]\label{prop: 2}
	Under Assumption \ref{assump: blinded},  $\tau_{\rm w}= \tau_{\rm wh} = \tau_{\rm whx} = E [ Y_i (1, \b{J}) ] - E [ Y_i (0, \b{J}) ]$.
\end{proposition}

Proposition \ref{prop: 2} shows that conditioning on  $\b{W}$ suffices to identify the treatment effect. This is because according to Assumption \ref{assump: blinded}, $\b{X}$ and  $\b{J}$ do not directly affect clusters' treatment assignment. However, adjusting for unit-level covariates and their aggregates may improve efficiency.

Note that target trial 2(a) is an instance of target trial 2 where the second stage randomization is independent of the first stage randomization. That is, the unit-cluster pairing does not depend on clusters' treatment assignments. In parallel, Assumption \ref{assump: blinded} and Proposition \ref{prop: 2} can also be applied to target trial 1 when the treatment assignment depends on clusters' characteristics $\b{W}$ but does not depend on unit-cluster pairing.  Therefore, we see that if the two stages of randomization are independent of each other, the identification conditions and results for target trial 1 and 2 are equivalent.	

In target trial 1 and 2(a) where the unit-cluster pairing does not depend on the treatment assignment, we can simplify the notation. That is, the index $\b{j}$ in the definition of potential outcomes can be omitted, since $\b{J}(1)= \b{J}(0)=\b{J}$. Specifically, we can define $Y_i(a_{J_i}): = Y_i(a_{J_i}, \b{J}) $ as the potential outcome for subject $i$. Under this notation, the possibility of units differentially selecting into clusters in response to treatment is precluded. Under this set of potential outcomes, the observed outcomes can be expressed as $Y_i = Y_i(A_{J_i}, \b{J}) = Y_i(A_{J_i}) =A_{J_i} Y_i(1) + (1- A_{J_i})Y_i(0)$, and the average treatment effect $E[Y_i(1, \b{J}) - Y_i(0, \b{J}) ]$ can be expressed in the following familiar form  $E[Y_i (1)- Y_i (0) ]$, and our identification result is consistent with the results in \cite{vanderweele2008ignorability}. This discussion highlights that extant work on identification in COS designs has implicitly assumed that $\b{J}$ is not affected by the treatment assignment and has ruled out the presence of differential selection \citep{hansen2014clustered,Keele:2015,Keele:2016b}. This assumption holds in target trial 1 where unit-cluster pairing precedes clusters' treatment assignment.  In target trial 2, this assumption is also innocuous when units are unaware that clusters have been assigned to the treatment. For example, checklists are often used to reduce medical errors. Hospitals may adopt such interventions with little to any awareness by the patients. In educational settings, many interventions may be alterations of the curriculum that students or parents are unaware of. Qualitative information on whether units are likely to be aware of the intervention will be critical to assessing the plausibility of target trial 2(a).

\subsubsection{Target Trial 2(b): Unblinded Unit-Cluster Pairing}

Now we consider the scenario under target trial 2 with unblinded unit-cluster pairing. As we noted above, we refer to this as target trial 2(b). Here, the unit-cluster pairing mediates the effect of treatment on the outcome.  Specifically, the unit-cluster pairing $\b{J}$ plays the role of a mediator that is affected by the treatment and also can directly affect subjects' outcomes; see review of mediation analysis in \cite{vanderweele2016mediation}. In this case, the primary causal effect of interest is arguably the direct effect of the treatment, because the goal is to learn about the effect of treatment itself separated from any effect due to changing the unit composition of the clusters.

To make progress under this target trial, we introduce an additional assumption about the structure of interference to further simplify the definition of potential outcomes, which may be plausible for a variety of settings. For any two $\b{j}$ and $\b{j}'$ such that $a_{j_i} = a_{j'_i}, W_{j_i} = W_{j'_i}$, and  $h( \b{X}_{[j_i]} )=h( \b{X}_{[j_i']} )$, we have $Y_i(a_{j_i}, \b{j}) = Y_i(a_{j'_i}, \b{j}') $.
This assumption intuitively asserts that the potential outcomes of subject $i$ remain the same as long as its associated cluster receives the same treatment, has the same cluster characteristics, and consists of subjects with the same individual covariate summaries. This type of assumptions is termed as stratified interference by \cite{hudgens2008toward}.  Under the stratified interference assumption, the potential outcome can be further simplified and written as $Y_i(a, w, h) =  Y_i(a_{j_i}=a, W_{j_i}=w,  h( \b{X}_{[j_i]} )=h ) $. Under target trial 2(b), the causal estimand is expressed as
\[
E [ Y_i (1, W_{J_i},  h(  \b{X}_{[J_i]} ) ) ]- E [ Y_i (0, W_{J_i},  h(  \b{X}_{[J_i]} ) ) ].
\]
\noindent This estimand is the average effect of the treatment for each unit while fixing the cluster- and aggregated individual-level characteristics of the associated cluster to the natural values that occur. As such, this estimand quantifies the effect that is purely due to the treatment. The stratified interference assumption renders this estimand equivalent to the estimand in \eqref{eq:estmd}.

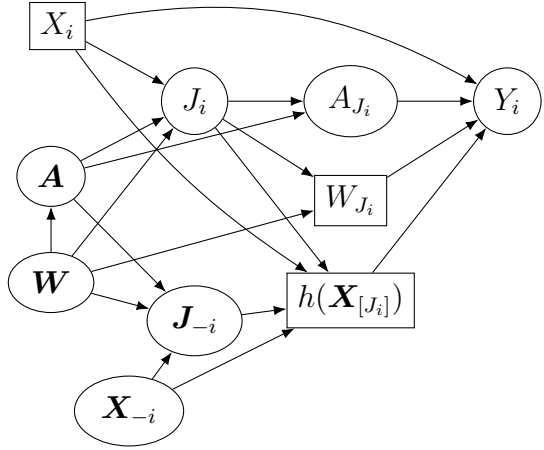
\begin{figure}
			\centering
			\centering
			\begin{tikzpicture}
				\node[state] (a) at (0,0) {${A}_{J_i}$};
				\node[state] (y) [right =of a] {$Y_i$};
				\node[state] (h) [rectangle, below =of a,yshift=-0.8cm] {$h(  \b{X}_{[J_i]} )$};
				\node[state] (ji) [left =of a] {$ J_i $};
				\node[state] (jo) [below =of ji, yshift=-1cm] {$ \b{J}_{-i} $};
				\node[state] (aall) [left =of ji, yshift=-1cm] {$ \b{A} $};
				\node[state] (w) [below= of aall, yshift=0.4cm] {$ \b{W} $};
				\node[state] (xi) [rectangle, left =of ji,yshift=1cm] {$ X_i $};
				\node[state] (xo) [left =of jo, xshift=1.5cm,yshift=-1.2cm] {$ \b{X}_{-i} $};
				\node[state] (wj) [rectangle, below= of a, yshift=0.5cm] {$ W_{J_i}$};
				\path (a) edge node[above]{} (y);
				\path (h) edge node[above]{} (y);
				\path (ji) edge node[above]{} (a);
				\path (aall) edge node[above]{} (ji);
				\path (aall) edge node[above]{} (jo);
				\path (aall) edge node[above]{} (a);
				\path (xi) edge node[above]{} (ji);
				\path (xo) edge node[above]{} (jo);
				\path (xo) edge node[above]{} (h);
				\path (jo) edge node[above]{} (h);
				\path (ji) edge node[above]{} (h);
				\path (xi) edge [bend right=-20]  (y);
				\path (w) edge (ji);
				\path (w) edge (jo);
				\path (w) edge (wj);
				\path (wj) edge (y);
				\path (ji) edge (wj);
				\path (w) edge (aall);
				\path (xi) edge  [bend right=10] (h);
			\end{tikzpicture}
			\caption{Directed acyclic graph (DAG) illustrating identification conditions for subject $i$ in target trial 2(b). Conditioning on $W_{J_i}, h( \b{X}_{[J_i]}), X_i $ renders the treatment-outcome relationship unconfounded.} \label{fig3}
		\end{figure}

Figure~\ref{fig3} contains the DAG for target trial 2(b). Unlike the DAGs for target trials 1 and 2(a), the DAG for target trial 3 is indexed by subject $i$ to account for unit selection to clusters, and there are now arrows from $ \b{A} $ to $ \b{J}_{-i} $ and $ J_i $ because of unblinded unit-cluster pairing.  Based on the DAG in Figure \ref{fig3}, we formalize the key assumption needed for identification using Assumption \ref{assump: unblinded}:
\begin{assumption} [Target Trial 2(b)]\label{assump: unblinded}
	$A_{J_i} \perp Y_i( a, w, h) \mid W_{J_i},  h(  \b{X}_{[J_i]} ), X_i $ for every $a, w, h, i$. 
		\end{assumption}
		
		
		Under this assumption, to disentangle treatment effect and peer effect, we propose comparing treated and control units that have similar observed characteristics and are in similar clusters with similar peers. The formal statement of identification under this assumption is contained in the following proposition:
		
		\begin{proposition}[Target Trial 2(b)] \label{prop: b}
			Under Assumption \ref{assump: unblinded}, $	\tau_{\rm whx} = E [ Y_i (1, W_{J_i},  h(  \b{X}_{[J_i]} ) ) ]- E [ Y_i (0, W_{J_i},  h(  \b{X}_{[J_i]} ) ) ].$
		\end{proposition}
		
		\noindent Proposition \ref{prop: b} shows that under target trial 2(b), the treatment effect is identifiable conditioning on  the cluster-level covariates $W_{J_i}$, individual-level covariates $X_i$, and aggregates of the individual-level covariates in the cluster $h( \b{X}_{[J_i]})$. Here, conditioning on $W_{J_i}, h( \b{X}_{[J_i]})$ is analogous to conditioning on mediators when the target parameter is the direct effect of the treatment \citep{vanderweele2016mediation}. Critically, we also need to adjust for $X_{i}$ to account for possible differential unit-level distributions within clusters. Therefore, unlike target trials 1 and 2, identification now depends on conditioning on the full set of baseline covariates.
		
The key insight from the identification results is that different target trials imply different adjustment sets (e.g., covariate sets on which we must condition). For target trial 1, identification is dependent on a conditioning set that includes $\{W, h\}$. For target trial 2(a), identification requires a conditioning on $\{W\}$, while target trial 2(b) requires conditioning on $\{ W, h, X\}$.  For applied analysts, knowing which target trial fits a given application will primarily depend on detailed knowledge of possible mechanisms for differential selection. That is, it is not possible to use the data to test between the different target trials. Critically, however, while it may not be possible to distinguish between the target trials in specific applications, we next use simulations to explore how using different conditioning sets can shed light on the appropriate target trial.

\subsection{Estimation and Inference}
		
If the appropriate adjustment set is selected, estimation is relatively straightforward. For example, to estimate $\tau_{\rm w}$, we can impose a parametric model of $\mu_{\rm w } (a, w)$, denoted by $\mu_{\rm w} (a, w; \eta_a) $. Let $\hat\eta_a$ denote the solution to the score equations corresponding to the likelihood of $Y_i$ conditional on $A_{J_i}=a $ and $W_{J_i}  $, the estimator of $\tau_{\rm w}$ is given by
		\[
		\hat \tau_{\rm w} = \frac1n \sum_{i=1}^n \mu_{\rm w}(1, W_{J_i} ; \hat\eta_1) - \frac1n \sum_{i=1}^n \mu_{\rm w}(0, W_{J_i};  \hat\eta_0) .
		\]
\noindent Estimators for $\hat \tau_{\rm wh}$ and $\hat \tau_{\rm whx}$ can be constructed in a similar fashion. This method of estimation is also called the parametric g-formula, see \citet[ch. 13]{Hernan-Robins} for a detailed review. Briefly, the estimation process consists of three steps.  First, we fit two outcome models, one for treated and one for control. Second, we obtain the fitted values for all $n$ subjects under the outcome models. Third, we standardize by separately averaging over the fitted values under treated and control, and calculate the difference. Variance estimators can be obtained using the block bootstrap which resamples both the clusters and all the units within the resampled clusters \citep{davison1997bootstrap}. We leave the development of more complex estimation methods to future work.
		
\section{Simulations}
		
		We conduct a simulation study to further evaluate how the specification of adjustment set can affect the amount of bias when estimating treatment effects in the COS design, especially when the adjustment set does not match the correct target trial. We consider  the following data-generating process. First, we generate the baseline covariates at the cluster- and unit- level in the population as:
		\begin{align*}
			&W_j\sim N(0,1), ~ j=1,\dots, m, \\
			&X_{i1} \sim N(0,1), ~ X_{i2} \sim {\rm Binom}(0.4), ~ i=1, \dots, n.
		\end{align*}
		For each target trial, we use these baseline covariates to govern how units are assigned to clusters. For target trial 1, we stipulate unit-cluster assignments with the following model:
		\begin{align}
			P(J_i = j\mid X_{i1}, X_{i2}, \bm W)= \frac{ \exp\{ 0.2W_j \cdot (1+X_{i1}+X_{i2})\} }{\left[ \sum_{j=1}^{m} \exp\{ 0.2W_j \cdot (1+X_{i1}+X_{i2})\} \right] } . \label{eq: unit-cluster}
		\end{align}
		Next, for a fixed unit-cluster pair, we calculate aggregate versions of $X_{i1}$ and $X_{i2}$. Specifically, we aggregate to the 25\%, 50\%, 75\% percentiles for $X_{i1}$'s and the mean for $X_{i2}$'s in cluster $j$, which are respectively denoted as $h_{j1}, h_{j2}, h_{j3}, h_{j4}$, for $j=1,\dots, m$. Finally, we assign clusters to treatment via the following model: $ {\rm logit} \{ P(A_j=1\mid W_j, h_{j1}, h_{j2}, h_{j3}, h_{j4}) \}= 0.2 W_j + 0.2(h_{j1} + h_{j2} + h_{j3}) + 0.2 (h_{j4} - 0.4 )$. Critically, consistent with target trial 1, treatment assignment depends on the cluster-level covariate and cluster aggregates of unit-level covariates.
		
		For target trial 2, we assign the treatment to clusters according to $ {\rm logit} \{ P(A_j=1\mid W_j ) \}= 0.2 W_j$, such that treatment assignment only depends on the cluster-level covariate, $ W_j $.  Then units are assigned to clusters using model \eqref{eq: unit-cluster} without information on clusters' treatment assignments. For target trial 3, we also assign the treatment according to $ {\rm logit} \{ P(A_j=1\mid W_j ) \}= 0.2 W_j  $, but we alter the model that assigns units to clusters using the following model:
		\begin{align*}
			P(J_i = j\mid X_{i1}, X_{i2}, \bm W)= \frac{ \exp\{ ( 0.2W_j + 0.2 A_j) \cdot (1+X_{i1}+X_{i2})\} }{\left[ \sum_{j=1}^{m} \exp\{  ( 0.2W_j + 0.2 A_j) \cdot (1+X_{i1}+X_{i2})\} \right] } .
		\end{align*}
The key difference is that unit to cluster assignment now depends on treatment assignment. For all three target trials, we generate outcomes using the same model:
		\begin{align*}
			Y_i = X_{i1} + X_{i2} + 0.4 A_{J_i} (X_{i1}+X_{i2}) + 0.5 (W_{J_i} + h_{J_i1} + h_{J_i2} + h_{J_i3} + h_{J_i4}) + 0.1 e_{J_i } + \epsilon_{i}
		\end{align*}
		where $e_j, \epsilon_{i}\sim N(0,1)$ for $j=1,\dots, m$, $i=1,\dots, n$. The true average treatment effect is $0.4 E[X_{i1} + X_{i2}] = 0.16$.

		The primary element of the analysis we vary is the adjustment set. That is, for each target trial, we use three different adjustment sets: $\{W\}, \{W, h\}$, and $\{ W, h, X\}$, through which we seek to understand how the adjustment set affects the treatment effect estimates. Specifically, we expect that under target trial 1, adjustment for $\{W, h\}$ should be sufficient for consistent estimation of the treatment effect. For target trial 2, adjustment for $\{W\}$ should be sufficient. For target trial 3, adjustment for $\{ W, h, X\}$ is necessary. We also vary sample sizes and consider $(m,n)=(50,4000), (100, 4000), $ and $(50, 8000)$. For each simulation scenario, we estimate treatment effects using separate linear model fits for the treated and untreated units. The standard errors are obtained using the block bootstrap with 300 bootstrap samples. We used 1,000 simulation repetitions for each scenario.
		
		\begin{table}[]
			\centering
			\caption{Mean, standard deviation (SD), average standard error (SE), and coverage probability (CP) of 95\% asymptotic confidence interval for the true average treatment effect $\tau =0.16$ based on 1,000 simulations. \label{tb:simu}}
			\resizebox{!}{0.58\textwidth}{\begin{tabular}{lllcccc} \toprule
				Target trial                & Sizes & Adjustment Set & Mean  & SD    & SE    & CP    \\  \hline
				1            & $m=50, n=4000 $              & $W$              & 0.185 & 0.106 & 0.104 & 0.929 \\
				&                            & $W, h$           & 0.160 & 0.051 & 0.080 & 0.974 \\
				&                            & $W, h, X$        & 0.160 & 0.049 & 0.078 & 0.966 \\ [1ex]
				& $m=100, n=4000 $              & $W$              & 0.199 & 0.102 & 0.099 & 0.916 \\
				&                            & $W, h$           & 0.158 & 0.043 & 0.045 & 0.954 \\
				&                            & $W, h, X$         & 0.158 & 0.040 & 0.042 & 0.946 \\[1ex]
				& $m=50, n=8000 $               & $W$              & 0.170 & 0.077 & 0.078 & 0.946 \\
				&                            & $W, h$           & 0.161 & 0.040 & 0.055 & 0.979 \\
				&                            & $W, h, X$        & 0.162 & 0.038 & 0.053 & 0.980 \\
				\midrule
				2   & $m=50, n=4000 $               & $W$              & 0.159 & 0.098 & 0.104 & 0.961 \\
				&                            & $W, h$           & 0.161 & 0.050 & 0.066 & 0.975 \\
				&                            & $W, h, X$         & 0.161 & 0.048 & 0.063 & 0.980 \\[1ex]
				& $m=100, n=4000 $              & $W$              & 0.160 & 0.099 & 0.099 & 0.943 \\
				&                            & $W, h$          & 0.160 & 0.043 & 0.044 & 0.952 \\
				&                            & $W, h, X$        & 0.160 & 0.041 & 0.042 & 0.947 \\[1ex]
				& $m=50, n=8000 $              & $W$              & 0.161 & 0.076 & 0.078 & 0.955 \\
				&                            & $W, h$           & 0.159 & 0.042 & 0.053 & 0.978 \\
				&                            & $W, h, X$        & 0.159 & 0.041 & 0.051 & 0.972 \\
				\midrule
				3 & $m=50, n=4000 $               & $W$              & 0.732 & 0.104 & 0.104 & 0.001 \\
				&                            & $W, h$           & 0.191 & 0.070 & 0.109 & 0.957 \\
				&                            & $W, h, X$        & 0.161 & 0.067 & 0.101 & 0.979 \\[1ex]
				& $m=100, n=4000 $              & $W$              & 0.739 & 0.097 & 0.099 & 0.000 \\
				&                            & $W, h$          & 0.188 & 0.052 & 0.054 & 0.926 \\
				&                            & $W, h, X$         & 0.160 & 0.050 & 0.052 & 0.962 \\[1ex]
				& $m=50, n=8000 $               & $W$              & 0.737 & 0.075 & 0.078 & 0.000 \\
				&                            & $W, h$           & 0.187 & 0.072 & 0.118 & 0.972 \\
				&                            & $W, h, X$         & 0.156 & 0.071 & 0.112 & 0.980 \\
				\bottomrule
			\end{tabular}}
		\end{table}

		Results from the simulation study are in Table~\ref{tb:simu}. First, we review the results for target trial 1. For target trial 1, if we only adjust for $\{W\}$ this leads to a biased estimates. However, the bias is relatively modest. When we only adjust for $\{W\}$, the average treatment effect is 0.19 versus the true treatment effect of 0.16 -- see row 1 of Table~\ref{tb:simu}. When we adjust for either $\{W,h\}$ or $\{ W, h, X\}$, the bias is negligible. For target trial 2, all three adjustment sets lead to unbiased treatment effect estimates, since all three adjustment sets contain $\{W\}$. For target trials 1 and 2, we find that doubling the number of clusters or doubling the total samples sizes substantially reduces the standard errors. However, in either case as long the adjustment set is appropriate, coverage rates perform as expected.
		
		For target trial 3, specifying the correct adjustment set is critical. Under target trial 3, if we only adjusting for $\{W\}$, the estimate is substantially biased. When we only adjust for $\{W\}$, the estimated treatment effect is too large by over a factor of 4. That is, the estimated treatment effects in this scenario are approximately 0.73 relative to the true treatment effect of 0.16. When we adjust for $\{W,h\}$, the bias is still present but much more modest with the estimate effect being approximately 0.19. Finally, if we adjust for the full set of covariates, $\{ W, h, X\}$, treatment effects are unbiased. For target trial 3, doubling the number of units does not lead to a reduction in the standard error estimates. That is, additional units do not increase the information in the data given the clustering of units. Here, doubling the number of clusters reduces the size of the standard errors and produces coverage probabilities that are close to the nominal level.
		
		The results from the simulation study agree with the identification results in Section~\ref{sec: identification}. Unbiased estimates depend critically on matching the correct adjustment set to the appropriate target trial. Critically, the key threat is from under-specification. Only under target trial 2 will adjustment for $\{W\}$ alone result in unbiased estimates. In general, we find that adjustment for the full set of covariates, $\{ W, h, X\}$, never leads to biased estimates. Given that in many applied applications there may be some uncertainty in terms of which target trial is appropriate, it would be wise to adjust for the full set of covariates.

		\section{Applications}
		
		We present two empirical applications: one from education and one from health services research, which are selected to contrast how key aspects of COS designs can vary depending on the applied context. For each application, we focus on the likelihood of differential selection of units to clusters. In both cases, differential selection is possible, but more likely in one case than in the other. In either case, we are unable to rule out the presence of differential selection.
		
		\subsection{Summer School Reading Intervention}
		
		In the first application, the empirical question of interest is whether a summer school reading intervention in Wake County, NC improved students' reading scores \citep{Keele:2016b,pagedesign2019}. Specifically, in the summer 2013, the Wake County Public School System (WCPSS) selected myON, a computer-aided reading program, for use in the summer school program for elementary school students. myON is a web-based software product designed to increase summer school attendees' reading comprehension. Due to technical constraints, only some summer-school sites used myON. WCPSS officials selected the schools that used myON, and principals and schools themselves had no input on program participation. Students at selected schools used the program for up to thirty minutes during the daily summer-school literacy block and could continue using it at home with a device and internet connection. Overall, 3,434 students from 49 different WCPSS elementary schools attended summer school. Of these, 1,371 summer-school students from 20 schools used myON. The primary outcome is student-level reading performance measured via standardized test scores.
		
		To begin, we consider which target trial is appropriate for analysis of the myON intervention. As this application illustrates, selecting the appropriate target trial analogue requires a detailed understanding of the substantive context. For myON, treatment assignment clearly is clustered in that the treatment was applied to all students in selected summer school sites. In general, we judge that the possibility of differential selection at the individual level is quite small. Specifically, students are selected for summer school based on their school-year performance and are residentially zoned into a particular summer school site. Further, summer school selection for myON occurred at the district offices. Students and parents were likely unaware of which summer school sites would be using myON. This is because myON was one relatively small part of the summer school curriculum, and the district did not advertise its use to students or parents. As such, we have no reason to believe that parents would have reacted to the selection of certain sites for myON by shifting student enrollment patterns in a way that would cause differential selection.  Nevertheless, we judge that either target trial 2(a) or 2(b) is more appropriate than target trial 1 because the selection of the summer school sites precedes students' summer school selection. 
		
It is important to note however, that we cannot rule out the possible presence of differential selection that would occur under target trial 2(b). Our evidence against differential selection is based on qualitative reasoning and not a statistical test. However, we can use balance tests for additional indirect evidence that target trial 2(a) holds. Table~\ref{tab:bal.bef} contains balance statistics for student- and school-level covariates. If differential selection did occur, we would expect student level covariates to be correlated with school level treatment assignment. For the myON application, we find that there are clear differences between treated and control schools in terms of school-level covariates such as proficiency in math and the share of teachers who are novices, but differences across student-level covariates are small.
		
		\begin{table}[ht]
			\centering
			\begin{threeparttable}
				\caption{Balance on student- and school-level covariates for myON application.}
				\label{tab:bal.bef}
				\begin{tabular}{lccc}
					\toprule
					Student Covariates & Treated Mean Before & Control Mean Before & Std. Difference \\
					\midrule
					Reading pretest score & 437.00 & 437.90 & -0.02 \\
					Math pretest score & 60.25 & 60.56 & -0.02 \\
					Male (0/1) & 0.36 & 0.40 & -0.09 \\
					Special education (0/1) & 0.47 & 0.43 & 0.09 \\
					Hispanic (0/1) & 0.53 & 0.52 & 0.02 \\
					African-American (0/1) & 0.22 & 0.22 & 0.00 \\
					\midrule
					School Covariates \\
					\midrule
					Composite proficiency & 60.74 & 58.56 & 0.21 \\
					Proficient in reading & 58.48 & 57.27 & 0.11 \\
					Proficient in math & 60.68 & 58.41 & 0.20 \\
					Free/reduced lunch eligible & 0.50 & 0.51 & -0.10 \\
					English language learners & 0.13 & 0.15 & -0.29 \\
					Novice teachers & 0.19 & 0.17 & 0.28 \\
					Staff turnover & 0.11 & 0.12 & -0.28 \\
					Nonwhite teachers & 0.14 & 0.18 & -0.26 \\
					Title I school & 0.90 & 0.93 & -0.11 \\
					\midrule
					Schools & 20 & 29 \\
					Summer school students & 1,371 & 2,063 & \\
					\bottomrule
				\end{tabular}
				\begin{tablenotes}[para]
					$Note$: Standardized difference for a given variable is computed as the mean difference between treatment and comparison schools or students divided by the pooled standard deviation.
				\end{tablenotes}
			\end{threeparttable}
		\end{table}
		
		Critically, our identification results do have observable implications for the role of confounders. Under target trial 2(a) to identify the effect of the myON intervention, we need to condition only on school-level covariates. Student-level covariates may improve efficiency of our estimates but are unnecessary for identification of the effect of interest. That is, if the assumptions of target trial 2(a) hold, we should not observe large differences in the magnitude of the point estimate across specifications that do and do not control for student-level covariates. In the analysis that follows, we consider specifications that include and omit student-level covariates.
		
		For this analysis, we estimate the myON treatment effect using the parametric g-formula based on a linear regression. We included quadratic terms for all continuous covariates. More flexible methods of estimation could be used to further expand the specification. We used 1,000 resamples from the block bootstrap to obtain Efron's percentile confidence intervals. For the specification that omits student-level covariates, the estimated treatment effect is 0.011, which implies that myON increased test scores 0.011 standard deviations. However, the confidence interval includes zero (95\% CI: -0.012, 0.034).  When we include student-level covariates, the estimated effect is 0.017 but the confidence interval is shorter (95\% CI: 0.003, 0.031). In sum, including student-level covariates increases the precision of our estimate slightly but does not  substantively change the magnitude, which provides further evidence that differential selection did not occur for the myON intervention.
		
		\subsection{Surgical Training}
		
		One strand of health services research focuses on whether certain aspects of surgical training have an effect on patient outcomes \citep{asch2009evaluating,bansal2016using,zaheer2017comparing,sullivan2012effect}.  Here, we re-analyze one study from this literature. \citet{sellers2018association} studied whether surgeons from university-based residency programs produce superior patient outcomes compared to surgeons trained in community-based residency programs. In their study, they used a data set that merges the American Medical Association (AMA) Physician Masterfile with all-payer hospital discharge claims from New York, Florida and Pennsylvania from 2012--2013. They collected data on residency type, and surgeons were classified as having attended either a university-based residency (UBR) or a non-university based residency (NUBR) based on the program listed in the AMA Masterfile. Data on surgeon age, sex and year of training completion were also collected. Surgeon experience was defined as year of training completion subtracted from year of operation. They compared surgeon performance between UBR and NUBR surgeons for patients that underwent one of 44 common operations performed by general surgeons in an inpatient setting. Operations were selected to capture a standard set of procedures routinely performed by general surgeons \citep{sellers2018association}. The data also contain patient sociodemographic and clinical characteristics including 31 comorbidities based on Elixhauser indices \citep{elixhauser1998comorbidity}. The primary outcome is a binary indicator of any postoperative complications that arise during the hospitalization. Complications were identified using ICD-9 diagnosis codes and collapsed into a binary variable indicating the development of 1 or more complications. For patients treated by UBR surgeons, 12.7\% had a post-operative complication. For patients treated by NUBR surgeons, 14\% had a post-operative complication. If we estimate the unadjusted treatment effect via a regression model with clustering at the surgeon level, the difference is statistically significant ($p$ = 0.001).
		
		The UBR study fits the COS template: all patients treated by a UBR surgeon are exposed to the treatment and vice-versa. However, the structure of the data for this application is quite different compared to the myON application. That is, the number of clusters and units is much larger. There are 498 treated surgeons and 1201 control surgeons. Overall, there are 86,305 patients operated on by UBR surgeons, and 193,307 patients operated on by NUBR surgeons. The number of patients treated by each surgeon varied from five to 1,074 over the two year period. Thus there are many more clusters, and there is considerable variation in the number of patients per surgeon. Moreover, we have 88 patient-level covariates but only five surgeon-level covariates.
		
		Next, we consider the possibility of differential selection. Unlike in the myON application, we have little qualitative evidence to rule out the possibility of differential selection. That is, it may be the case that if UBR surgeons are viewed as more skilled, they will be assigned patients that have more complex pre-operative conditions or with a generally worse prognosis. As such, differential selection is an open possibility in this application. Again, we can use balance statistics to shed light on this possibility. Specifically, we consider whether UBR patients are observably different from NUBR patients by examining standardized mean differences between patients of NUMBER and UBR surgeons. Surprisingly, we found that none of the patient-level covariates had standardized differences larger than 0.10. This suggests that differential selection may not be in operation. Still, we cannot rule out that UBR patients have higher levels of unobserved frailty.
		
		Next, we estimate the UBR effect using the parametric g-formula via linear regression. In our analysis, we used three different specifications. The first specification only controls for surgeon-level variables. This specification is consistent with target trial 1. In the second specification, in addition to the surgeon-level covariates we included the patient-level variables aggregated to the surgeon level. For the aggregation, we used the average. This specification would identify the average treatment effect under the assumptions of target trial 2(a). In the third specification, we include patient-level variables as well. This specification would identify the average treatment effect under the assumptions of target trial 2(b). We used 1,000 resamples from a block bootstrap to obtain Efron's percentile confidence intervals.
			
		For the surgeon only specification, UBR surgeons had lower complications by 1.5 percentage points (95\% CI: -0.022, -0.008). Thus controlling for surgeon level covariates leaves the estimate nearly unchanged compared to the unadjusted estimate. For the second specification, the difference between UBR and NUBR surgeons is -0.008 (95\% CI: -0.014, -0.003). Now the estimated difference in complication rates is quite small. As such, adding the aggregated patient covariates reduces the magnitude of the treatment effect substantially. For the third specification, the difference between UBR and NUBR surgeons is -0.006 (95\% CI: -0.010, -0.002). 	As such, adding patient level covariates does not change the estimated effect. This suggests that in this application, patient assignment to surgeons is not a function of the training, but instead follows some unrelated cluster level mechanism.	
		
		\section{Discussion}
		
		The literature on the COS design has primarily focused on estimation methods and has operated under identification assumptions that are essentially borrowed directly from study designs with unclustered treatment assignment. However, those identification assumptions imply that differential selection does not occur. That is, prior work has assumed that when treatments are assigned to clusters, this assignment has no effect on the population of units within the clusters. While this assumption may be innocuous in some settings, it is implausible in others. For example, in settings where the treatment is a school-wide reform effort, such as Success for All \citep{borman2007final}, parents may react and move to or away from the schools that are exposed to the treatment. In this paper, we consider how differential selection changes the identification assumptions in the context of COSs.
		
		We used the target trial framework to formalize identification conditions for the COS design with and without differential selection. We outlined three possible target trials to describe different scenarios for both the assignment of treatments and units to clusters. In this framework, we show that for each target trial, the set of covariates that render treatment assignment ignorable differ.   Under target trial 1, analysts need to condition on cluster-level covariates and cluster-level aggregates. Under target trial 2, analysts need to condition on cluster-level covariates. Only under target trial 3, do investigators need to condition on cluster covariates, cluster aggregates, and unit-level covariates. For target trial 3, it is critical to condition on unit-level covariates to control for the possible differential mix in cluster populations.
		
		Our work has key implications for applied research. In many cases, researchers may not have definitive information on whether differential selection is present in a COS design. As we demonstrated in our empirical examples, we may be able to reason about the likelihood of differential selection but still be unable to rule it out. Critically, we show that conditioning on the full set of covariates will reduce bias if differential selection is present.  However, conditioning on the full set of covariates does no harm if differential selection did not occur. As such, researchers should consider more expansive specifications to reduce possible bias from differential selection. As an alternative, investigators can compare a specification that omits unit-level covariates and a specification that includes unit-level covariates.  If the magnitude of the treatment effect estimate differs across these two specifications, it  provides some evidence for differential selection. As we highlighted, COS designs are common in many areas of applied research. Our work provides researchers in these areas with guidance on how to consider the possibility of differential selection and  how to select specifications that reduce bias.

		\clearpage
		
		\bibliographystyle{apalike}
		\bibliography{cluster}

		\clearpage
		
		\setcounter{equation}{0}
		\setcounter{table}{0}
		\setcounter{lemma}{0}
		\setcounter{section}{0}
		\setcounter{figure}{0}
		\setcounter{theorem}{0}
		\renewcommand{\theequation}{S\arabic{equation}}
		\renewcommand{\thelemma}{S\arabic{lemma}}
		\renewcommand{\thetheorem}{S\arabic{theorem}}
		\renewcommand{\thefigure}{S\arabic{figure}}
		\renewcommand{\thesection}{S\arabic{section}}

		\begin{center}
			\LARGE \bf Supplementary Material
		\end{center}
		\section{Technical Proofs}
		\subsection{Proof of Proposition \ref{prop: a}}
		First note	\begin{align*}
			&E[ Y_i (a, \b{J}) \mid  A_{J_i} = a, W_{J_i}=w,
			h(  \b{X}_{[J_i]} ) = h , X_i = x  ]  \\
			&=	\sum_{j=1}^m E[ Y_i (a, \b{J}) \mid A_{j} = a,   W_{j}=w,
			h( \b{X}_{[j]} ) = h , X_i = x , J_i = j]  \\
			&\qquad\qquad P(J_i = j\mid  A_{J_i} = a, W_{J_i}=w,
			h(  \b{X}_{[J_i]} ) = h , X_i = x)  \\
			&= 		\sum_{j=1}^m E[ Y_i (a, \b{J}) \mid W_{j}=w,
			h(  \b{X}_{[j]} ) = h , X_i = x , J_i = j]  P(J_i = j\mid  A_{J_i} = a, W_{J_i}=w,
			h(  \b{X}_{[J_i]} ) = h , X_i = x)  \\
			&= 		\sum_{j=1}^m E[ Y_i (a, \b{J}) \mid W_{J_i}=w,
			h(  \b{X}_{[J_i]} ) = h , X_i = x , J_i = j]  P(J_i = j \mid  W_{J_i}=w,
			h(  \b{X}_{[J_i]} ) = h , X_i = x)  \\
			&= 	E[ Y_i (a, \b{J}) \mid W_{J_i}=w,
			h(  \b{X}_{[J_i]} ) = h , X_i = x ]
		\end{align*}  	
		where the second equality is because Assumption \ref{assump: pairing first} implies that  $A_j \perp \{ X_i,Y_i (a, \b{J}), J_i
		\} \mid W_j,  h(  \b{X}_{[j]} )$ for every $i, j$, and thus $A_j \perp Y_i (a, \b{J})
		\mid W_j,  h(  \b{X}_{[j]} ), X_i , J_i $ for every $i, j$, and the third equality is because $A_{J_i}\perp J_i \mid W_{J_i}, 	h(  \b{X}_{[J_i]} ), X_i $ from $P(A_{J_i} = 1\mid J_i=j,W_{J_i} = w, 	h(  \b{X}_{[J_i]} ) = h , X_i = x ) = P(A_{j} = 1\mid W_{j} = w, 	h(  \b{X}_{[j]} ) = h, X_i= x  )  = \pi_A(w, h,x)=  P(A_{J_i} = 1\mid W_{J_i} = w, 	h(  \b{X}_{[J_i]} ) = h, X_i= x  ) $ also from Assumption \ref{assump: pairing first}, where $\pi_A(w, h,x):= P(A_{j} = 1\mid W_{j} = w, 	h(  \b{X}_{[j]} ) = h, X_i= x)$. Then,
		\begin{align*}
			& E[ Y_i \mid  A_{J_i }=a, W_{J_i}=w,
			h(  \b{X}_{[J_i]} )= h, X_i = x ] \\
			&= E[ Y_i (a, \b{J}) \mid  A_{J_i }=a, W_{J_i}=w,
			h(  \b{X}_{[J_i]} )= h, X_i = x ] \\	
			&= E[ Y_i (a, \b{J}) \mid  W_{J_i}=w,
			h(  \b{X}_{[J_i]} ) = h , X_i = x ] .
		\end{align*}
		Hence, $E[ \mu_{\rm whx}(a, W_{J_i},  h(  \b{X}_{[J_i]} ), X_i) ] = E [ Y_i (a, \b{J}) ]$.
		The above results also hold without conditioning on $X_i$ and thus $E[ \mu_{\rm wh}(a, W_{J_i},  h(  \b{X}_{[J_i]} )) ] = E [ Y_i (a, \b{J}) ]$.

		\subsection{Proof of Proposition \ref{prop: 2}}
		
		First note that Assumption \ref{assump: blinded} implies  Assumption \ref{assump: pairing first}. Hence the results proved in Proposition \ref{prop: a} still holds under  Assumption \ref{assump: blinded}. The results when conditioning on $W_{J_i}$ can be proved in the same way.

		\subsection{Proof of Proposition \ref{prop: b}}
		From Assumption \ref{prop: b},
		\begin{align*}
			& E[ Y_i \mid  A_{J_i }=a, W_{J_i}=w,
			h(  \b{X}_{[J_i]} ) = h, X_i = x ] \\
			&= E[ Y_i (a, w, h ) \mid  A_{J_i }=a, W_{J_i}=w,
			h(  \b{X}_{[J_i]} )= h, X_i = x ] \\	
			&= E[ Y_i (a, w, h )\mid  W_{J_i}=w,
			h(  \b{X}_{[J_i]} )= h , X_i = x ] ,
		\end{align*}
		Hence, $E[ \mu_{\rm whx}(a, W_{J_i}, h(  \b{X}_{[J_i]} ),  X_i) ] = E [ Y_i (a, W_{J_i},  h(  \b{X}_{[J_i]} ) ) ]$.

\end{document}